\documentclass{aa}  

\usepackage{txfonts}
\usepackage{graphicx}	
\usepackage{amsmath}	
\usepackage{soul}
\usepackage{xspace}
\usepackage{subcaption}
\usepackage{gensymb}
\usepackage{bm}	
\usepackage[T1]{fontenc}
\usepackage{xcolor}
\newcommand\bluesout{\bgroup\markoverwith{\textcolor{blue}{\rule[0.5ex]{2pt}{0.4pt}}}\ULon}
\usepackage[normalem]{ulem} 
\definecolor{lightblue}{rgb}{0.1,0.5,0.89}

\definecolor{xlinkcolor}{cmyk}{1,1,0,0}
\usepackage[
  bookmarks=true,         
  pdfnewwindow=true,      
  colorlinks=true,        
  linkcolor=xlinkcolor,   
  citecolor=xlinkcolor,   
  filecolor=xlinkcolor,   
  urlcolor=xlinkcolor,    
  final=true,
]{hyperref}
\usepackage[capitalize]{cleveref}
\begin{document}

\title{Barren but not Empty: The Impact of Void Environments on Galaxy and Halo Populations}
\titlerunning{The Halo Mass Function in Voids}

   \author{Anita M. Schiller
          \inst{\ref{inst:usm}}
          \and
          Benjamin A. Seidel\inst{\ref{inst:usm}}
          \and
           Rhea-Silvia Remus\inst{\ref{inst:usm},\ref{inst:swin}}
           \and
           Lucas C. Kimmig\inst{\ref{inst:usm},\ref{inst:nott}}
           \and
           Klaus Dolag\inst{\ref{inst:usm},\ref{inst:mpa}}
          }

   \institute{Universitäts-Sternwarte, Fakultät für Physik, Ludwig-Maximilians Universität München, Scheinerstr.1, 81679 München, Germany\label{inst:usm}
   \and 
   Centre for Astrophysics and Supercomputing, Swinburne University of Technology, Hawthorn VIC 3122, Australia\label{inst:swin}
   \and
   School of Physics and Astronomy, University of Nottingham, University Park, Nottingham NG7 2RD, UK\label{inst:nott}
   \and 
   Max-Planck-Institute for Astrophysics, Karl-Schwarzschild-Str.\ 1, 85748 Garching, Germany\label{inst:mpa}
             }

   \date{Received August 26, 2026; accepted ***}


\abstract
{The combination of gravitational collapse and an accelerating expansion of the Universe drive an increasing fraction of volume toward becoming low-density voids across cosmic time. These regions present a unique environment due to their low density, which leads to them behaving similarly to "pocket universes" with modified cosmological parameters. Since the formation and evolution of galaxies is strongly tied to the environment and cosmological conditions, we aim to discern how these void environments alter the halo and galaxy population when compared to the general population in the universe. 
We use the fully hydrodynamical cosmological simulation suite Magneticum Pathfinder and introduce a set of void halo mass functions (VHMFs) which are based on enclosed spherical shells from the void centers, tracing the halo population as a function of void-centric distance. As expected, we find galactic halos in voids to be overall less massive than in the general simulation volume. This discrepancy between void and field halos evolves with time as the voids grow more underdense, while the populations are still very similar at cosmic dawn.
Expanding on this, we evaluate whether specific properties of the voids affect the VHMFs. We find the only relevant parameter for the void halo population to be the core density, a measure of the strength of underdensity. To assess the impact of the density environment on the hydrodynamical evolution of these halos, we investigate whether there are any distinctions between void and field halos with regard to stellar properties of their galaxies. We compute the stellar-mass function, halo-mass-stellar-mass relation, as well as the star-forming main sequence using the same void-centric shells. Although the voids contain overall less stellar mass, the distribution of galaxies along the stellar mass-halo mass actually follows that of other regions. This implies that SF proceeded "conventionally" in void regions, in agreement with more recent observations. This similar growth of void and field galaxies holds until the massive end ($M_*>10^{10.5} M_\odot$), at which point the void galaxies do exhibit a reduction in SF rate at late cosmic times.}

   
   
  
 


   \maketitle
%

\section{Introduction}
The large-scale structure of our universe has long been known to be a web-like structure: galaxies assemble into thin filaments and sheets, with the densest structures, galaxy clusters, at the intersecting nodes of these filaments.\citep{zeldovich1970b}. The space between these filaments and nodes is largely vacant; these underdense regions of space have been dubbed as "voids" \citep[e.g.,][]{bond:1996}. Whether these different types of environments have an impact on the galaxy populations they host is currently under investigation, especially since these environments themselves change strongly with the evolution of the Universe. 

One of the fundamental scaling relations in our Universe is the halo mass function (HMF), closely related to the statistical properties of the underlying density field set by the cosmology, as well as the distribution of halos and their galaxies throughout the Universe and over cosmic time. Theoretical models \citep[e.g.][]{press:1974,schechter:1976,peebles:1993} have been employed to explain the existence of this halo mass function, and extensive comparisons with simulations and observations have led us to believe that the Universe, in fact, is close to a $\Lambda$-Cold-Dark-Matter Universe \citep{riess1998a,beutler2017,planckcollaboration2016}. 
In principle, the halo mass function includes the halo population of both underdense and overdense regions. However, when estimating the halo mass function from a given cosmological volume - as both simulations and observations usually do - this statistic is heavily dominated by overdense regions such as clusters and filaments. This effect is even larger when the halo mass function is estimated observationally from the surroundings of individual galaxy clusters \citep[e.g.][]{valageas2009}.

The shape and nature of the halo mass function inside voids, the void halo mass function (VHMF), is largely unstudied, leaving the question unanswered: are voids populated by the same kinds of halos as filaments and nodes, and does the void environment change the mass distribution of halos? 
From a dark-matter-only simulation suite and a theoretical modeling approach, \citet{verza:2022} found a dependence of the halo mass function in voids on the radial distance to the void center, with less massive halos being present the further inside the void the halo mass function is calculated. From a purely theoretical side, this trend of halos within the voids being less massive is predicted with excursion set theory by deriving statistics for the halo-in-void trajectories within a two-threshold framework, though there is no direct prediction for radial trends in this formalism \citep[e.g.,][]{song:2009,parkavousi:2023}. Complicating observational verification of this effect is the fact that the halo mass function is difficult to measure directly from observations, as the stellar luminosity function is much easier to obtain, which convolves pure mass trends with the star formation physics. \citet{moorman:2015} report, using SDSS and ALFALFA data, that the low luminosity end of the luminosity function in voids agrees well with the luminosity function found in filaments and walls, while the higher mass end is less populated and the characteristic magnitude is shifted towards lower magnitudes. Similar results were previously found by \citet{croton:2005} using the 2dF Galaxy Redshift Survey; however, while they report that the luminosity function can be well described by a Schechter function \citep{schechter:1976}, \citet{moorman:2015} do not find this for the higher-mass end.

Observationally, the existence of galaxies in voids has been well established; however, whether the population differs from that in filaments and walls is still under debate. 
On the one hand, indications have been reported for void galaxies to be more disk-like in structure, color, and star formation rates \citep[e.g.,][]{szomoru:1996,rojas:2004,kreckel:2012,hoyle:2012,beygu:2017,porter:2023}, as well as generally less massive \citep[e.g.,][]{moorman:2015}. Their specific star formation rates are overall found to be higher than those reported for the filaments and walls \citep[e.g.,][]{rojas:2005,moorman:2016, beygu:2016,florenz:2021}, and indications from dwarf galaxies are found that the individual metal abundances are different in the voids compared to the field \citep{douglass:2018} with overall lower N/O ratios possibly indicating delayed star formation in these evironments.
In addition, the AGN fraction in the inner regions of voids is reported to be heightened, compared not only to the outer regions of voids \citep{mishra:2021} but also to the field \citep{ceccarelli:2022}, indicating that the void region actually enhances the activity of AGN.
On the other hand, other studies did not find differences between the galaxy populations in voids and those in filaments or walls, particularly in star formation rates and masses \citep[e.g.,][]{patiri:2006,ricciardelli:2014}. In particular, at the high stellar mass end, \citet{fraser-mckelvie:2016} reported that the void galaxies have similarly low star formation rates as those galaxies of similar mass that live in the field.

Recently, the large, systematic integral-field unit study, the CAVITY survey \citep{perez:2024}, added a more spatially resolved and statistically representative sample of void galaxies to the discussion. First results from this study indicate that void galaxies have lower metallicities at a given stellar mass than their counterparts in filaments and especially galaxy clusters \citep{dominguezGomez:2023}, and are overall larger and younger \citep{conrado:2024}, indicative of a more gradual, slower evolution. Their star formation histories in particular show that they have assembled more slowly than their non-void counterparts \citep{dominguesGomez:2023Natur}. 
However, \citet{conrado:2024} do not find a difference in the quenched fraction of galaxies inside voids when compared to the field following the criterion for quiescence as applied by \citet{bluck:2016}, in contradiction to what has been reported by \citet{kuutma:2017} from SDSS. In addition, regarding their molecular gas content, the ALMA CO-CAVITY survey did not find systematic differences between the wall and filament counterparts \citep{dominguesGomez:2022}.

From a simulation perspective, studying the properties of void galaxies in a statistically meaningful way is difficult, as it requires a large simulation volume to obtain a sufficient number of voids and their galaxies, while also providing sufficient resolution to study individual galaxy properties. Such simulations do not currently exist, as all simulations with well-resolved galaxies have box volumes of $(100~\mathrm{Mpc}/h)^3$ or less, which contain only a handful of voids at $z=0$. Nevertheless, despite being limited in void numbers, interesting results have emerged from such simulations in recent years: \citet{rosasGuevara:2022} report using the Eagle cosmological simulations that the star formation activity and the atomic hydrogen mass are enhanced for galaxies inside the voids, and they find the quenched fractions to be lower in voids for galaxies below $\log M_*<9.5$, and higher for galaxies with $9.5 < \log M_* < 10.5$ compared to denser environments. For galaxies above $\log M_*>10.5$, they do not find any differences to other environments. Using the HorizonAGN simulations, \citet{habouzit:2020} find similar trends in terms of void galaxies being overall less massive; however, they find that void galaxies in general form stars more efficiently than their non-void counterparts. While galaxies in voids have predominantly low-mass black holes, they resemble their non-void counterparts of the same stellar mass in terms of black hole mass, showing no differences beyond the low-mass galaxy prevalence. Both simulations agree with some observations, though they cannot confirm the others. Unfortunately, these simulations are too small to study the possible effects of void properties on the results, which could potentially be a source of the differences in observations.
Those simulations with larger volumes are not well enough resolved to study the properties of individual galaxies in detail; however, they can provide some basic answers to the global properties of the galaxies inside the voids in a more statistical manner. Using the IllustrisTNG-300 simulations, \citet{rodriguezMedrano:2024} find generally higher star formation rates, higher gas metallicities, and lower stellar metallicities for galaxies in voids, but also surprisingly a larger amount of accreted stars in low-mass void galaxies compared to their non-void counterparts, in agreement with the other simulations. In addition, \citet{alfaro:2020} find void halos to be younger both from the IllustrisTNG-300 simulations as well as the SAG semi-analytical model. 

Interestingly, all simulations agree in that they find void galaxies to have systematically smaller stellar masses at a given halo mass \citep[i.e.,][]{habouzit:2020,alfaro:2020,rosasGuevara:2022,rodriguezMedrano:2024}, a result that is further supported by zoom-simulations on voids as reported by \citet{rodriguesMedrano:2022}. However, this is in disagreement with observational results from \citet{douglas:2019}, using SDSS MaNGA, who do not find any difference in stellar-to-halo mass for void galaxies and non-void galaxies using H$\alpha$ as tracers of the dark matter. The origin of this discrepancy should be studied in more detail in the future.

In this work, we will employ one of the largest cosmological simulations with sufficient resolution to study global galaxy properties from the Magneticum Pathfinder simulations, Box2b with a box volume of $(640~\mathrm{Mpc}/h)^{3}$, to address the question of whether the halo mass function inside voids differs from the halo mass function seen globally, and whether such differences might be correlated to the properties of the host-voids, as this simulation volume is large enough to statistically account for such void-intrinsic differences. We will also examine the evolution of the VHMF with redshift and void radius. This paper is structured as follows: In Section~\ref{sec:sim}, the simulation is introduced along with the method for identifying voids and the galaxies within them. In Section~\ref{sec:results}, we will first discuss the global properties of the different voids. Then we will study how the void halo mass functions depend on the void properties and simulation resolution at varying void radii through cosmic time. Furthermore, we investigate the impact of voids on the baryonic processes of galaxies of masses larger than $M_* = 10^{10} M_{\odot}$ on a statistical scale. In Section~\ref{sec:conclude}, we summarize and discuss the results.

\section{Simulation and Void Finding Procedure}\label{sec:sim}
As the primary objective of this study is to understand the population of halos within voids, we require a simulation large enough to contain a statistically relevant number of voids while also being resolved well enough to include halos down to about Milky Way mass. In addition, halos and voids need to be identified. In the following, we first introduce the simulation suite used in this study and the method for identifying voids and obtaining their properties.

\subsection{The Magneticum Pathfinder Simulations}\label{sec:mag}
The \textit{Magneticum Pathfinder}\footnote{www.magneticum.org} simulations \citep{hirschmann:2014, teklu:2015, dolag2025} are a suite of hydrodynamical cosmological simulations covering a large range of box volumes, ranging from $(2688~\mathrm{Mpc}/h)^3$ to $(48~\mathrm{Mpc}/h)^3$, with different levels of resolutions. They were performed with an updated version of GADGET-3 based on GADGET-2 by \citet{springel:2005}. These improvements include updates to the formulation of SPH according to \citep{dolag:2004,dolag:2005,donnert:2013,beck:2015} and the inclusion of thermal conduction \citep{dolag:2004}, including a Spitzer-value of 1/20 following \citet{arth:2014}. The subgrid physics are, in a nutshell, implemented as follows: star formation and metal enrichment follow \citet{tornatore:2004,tornatore:2007} and \citet{wiersma:2009}, see also \citet{dolag:2017} for more details on the metal yields. The AGN feedback is implemented following \citet{fabjan:2010}, with more details on the model described by \citet{hirschmann:2014} and \citet{teklu:2015}. 
\textit{Magneticum Pathfinder} employs a WMAP-7 $\Lambda$CDM cosmology \citep{komatsu:2011}, with the cosmological parameters chosen as $\sigma_8 =0.809$, $h = 0.704$, $\Omega_\Lambda = 0.728$, $\Omega_\mathrm{M} = 0.272$, $\Omega_\mathrm{B} = 0.0451$ and $n_\mathrm{s} = 0.963$. 

Throughout this work, we use two different box volumes with different resolutions. The main part of the work is performed using {\it Box2b/hr}, which has a box size of $(640~\mathrm{Mpc}/h)^{3}$, with initially $2\times2880^{3}$ (dark matter and gas) particles. The simulation has a mass resolution of $m_\mathrm{dm} = 6.9\times10^{8} M_{\odot}/h$ and $m_\mathrm{gas} = 1.4\times10^{8} M_{\odot}/h$ for dark matter and gas particles, respectively, and every gas particle can spawn up to four stellar particles, resulting in an average stellar particle mass of $m_\mathrm{*}\approx 3.5\times10^{7} M_{\odot}/h$. The softening length is $\epsilon_\mathrm{dm} = \epsilon_\mathrm{gas} = 3.75~\mathrm{kpc}/h$ for dark matter and gas particles, and $\epsilon_\mathrm{*} = 2~\mathrm{kpc}/h$ for stellar particles. This box volume is chosen for its large size, which includes a large number of voids as we will discuss in Section~\ref{sec:vide}. This simulation has been shown to successfully trace halo evolution through cosmic time while reproducing observed properties \citep[e.g.,][]{lotz:2019,remus:2023,kimmig:2023}.

In addition, we include results obtained from a smaller but higher resolved simulation, namely \textit{Box3/uhr}, with a box size of $(128~\mathrm{Mpc}/h)^{3}$, with initially $2\times1536^{3}$ (dark matter and gas) particles. The simulation has a mass resolution of $m_\mathrm{dm} = 3.6\times10^{7} M_{\odot}/h$, $m_\mathrm{gas} = 7.3\times10^{6} M_{\odot}/h$, and $m_\mathrm{*}\approx 2\times10^{6} M_{\odot}/h$. The softening length is $\epsilon_\mathrm{dm} = \epsilon_\mathrm{gas} = 1.4~\mathrm{kpc}/h$ for dark matter and gas particles, and $\epsilon_\mathrm{*} = 0.7~\mathrm{kpc}/h$ for stellar particles.
This box volume is much smaller than \textit{Box2b/hr}, but still large enough to contain a reasonable amount of voids. Due to the higher resolution, halos can be found down to lower masses than in \textit{Box2b/hr}, thereby complementing the larger statistics from \textit{Box2b/hr} at the low-mass end and serving as a resolution study. However, this simulation only ran down to $z=1.9$, so comparisons are limited to the high redshift range. In this redshift range, the properties of early forming galaxies have been shown to agree well with the latest high-redshift observations in terms of their stellar masses and general properties \citep{kimmig2025a}.

Halos and their galaxies are identified in the simulations using a modified baryonic version of SUBFIND \citep{springel:2001,dolag:2009}. For \textit{Box2b/hr}, halos with masses larger than $M_\mathrm{vir}>5\times10^{10}M_\odot/h$ can be identified and analyzed, while for \textit{Box3/uhr} halos with masses above $M_\mathrm{vir}>5\times10^{9}M_\odot/h$ are included.

\subsection{The Void Finding Algorithm VIDE}\label{sec:vide}
To extract the voids from the simulation, we use the VIDE void finding algorithm \citep{sutter2014}, based on the ZOBOV void finder \citep{neyrinck2008}. Previous work applying VIDE to the Magneticum simulations demonstrated that the voids obtained with this approach follow the predictions of linear theory very well by examining their velocity and density profiles for different resolutions and tracer particles 
\citep{schuster:2024}.

VIDE is a shape-free void finder where voids are identified by computing a Voronoi tessellation on the tracers and grouping underdense cells to their local minima. Either dark matter particles or halos can be chosen as tracers; in this work, we use dark matter particles exclusively as unbiased tracers of the true density field. We additionally apply a down-sampling procedure to the dark matter particles, reducing the total particle number to a desired degree by random sampling. To sample similar voids in different boxes, we keep the mean tracer separation constant at a value of $n=4\times10^{-3} ({\mathrm{Mpc}/h})^{-3}$. This approach emulates the sparse spatial sampling of the density field when using galaxies as tracers, while avoiding the halo bias. The spatial tracer density is comparable to the lowest sampling densities used in \citet{schuster:2024} in order to probe the largest features in the density field. The regions created this way can then be further merged by performing a \textit{Watershed} transform, with the maximum ridge density that can be merged over a free parameter chosen by the user. Unless specified otherwise, the voids in this study are selected from the unmerged (or children) population.

The VIDE output provides for each void a center position and an effective radius $r_v$ based on the sum over the volumes of its constituent Voronoi cells. These outputs are the basis for our void-halo catalog (see sec \ref{sec:results}). An Additional void property used in this work (see \cref{sec:results}) is their core density, defined as $\rho_{void}/{\bar{\rho}_{box}}$, the ratio of the mean density of the simulation volume to the lowest density in the void center. This is a simple proxy for the steepness of the density profile in a void.
A similar proxy for the shape of a given void can be calculated with the output eigenvalues of the void's inertia tensor via
\begin{equation}
    \epsilon=1-{\left(\frac{J_2+J_3-J_1}{J_1+J_2-J_3}\right)}^{1/2}
\end{equation}
where $J_1>J_2>J_3$ are the ordered eigenvalues. It should be noted that this definition deviates from the standard VIDE output ellipticity ($\epsilon=1-{(\frac{J_1}{J_3})}^{1/4}$). We chose this definition because it is closer to analytical descriptions of tidal-field statistics and void shapes (see Seidel et al. in prep.).

\section{The Void Halo Mass Function through Cosmic Time}\label{sec:results}
\begin{figure*}
   \centering
   \includegraphics[width=0.95\textwidth]{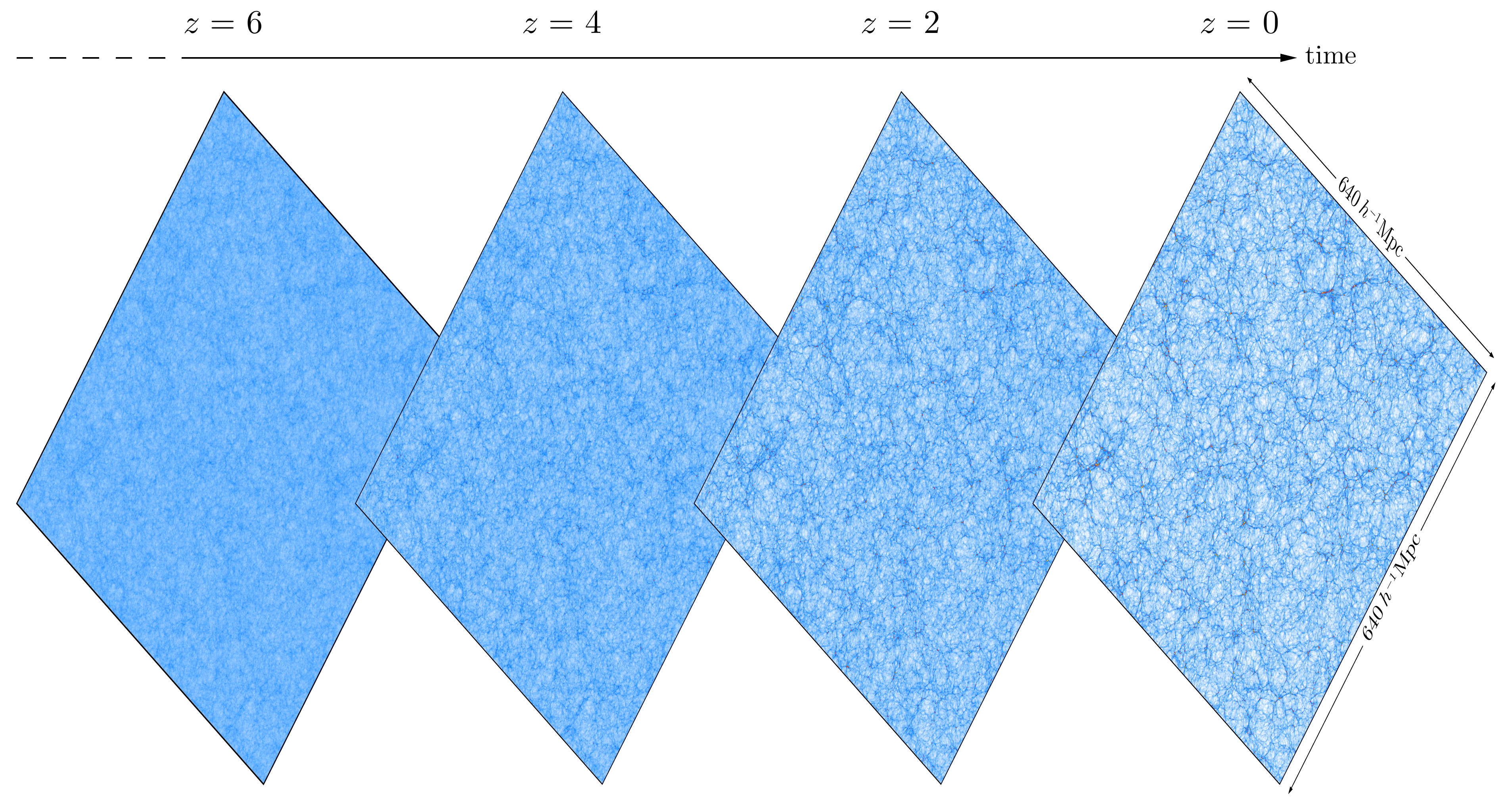}
   \caption{Evolution of the cosmic web structure in slices of $2~\textrm{Mpc/h}$ thickness through \textit{Box2b}, at redshifts of $z=6.9$, $z=4.2$, $z=1.9$, and $z=0.25$ from left to right. The color shows the dark matter density of the cosmic web, with empty voids in white, the web shown in blue, and the most massive nodes appearing in red. 
   }
   \label{fig:boxview}%
\end{figure*}
Using VIDE, we identified $6443$ voids at $z=0.25$, $9864$ voids at $z=1.9$, $12195$ voids at $z=4.2$, and $13026$ voids at $z=6.9$ in \textit{Box2b}. In the higher resolution \textit{Box3} we identified $78$ voids at $z=1.9$, $112$ voids at $z=4.2$, and $101$ voids at $z=6.9$.
\begin{table}[h]
    \centering
    \begin{tabular}{c| c c c c}
         Redshift & 6.9 & 4.2 & 1.9 & 0.25   \\
         \hline \hline 
         \textit{Box2b} & 13026 & 12195 & 9864 & 6443  \\
         \textit{Box3} & 101 & 112  & 78 & - \\
    \end{tabular}
    \caption{Number of voids in the different simulation volumes \textit{Box2b} and \textit{Box3}}
    \label{tab:voidcounts}
\end{table}
This already indicates what we also see in Fig.~\ref{fig:boxview}, namely that the cosmic web is initially fairly indistinct at high redshifts and grows increasingly pronounced at lower redshifts, which leads to the voids appearing successively more distinguished with lower redshift. In addition, we see that the total void number is largest at $z=6.9$, with void mergers and the vanishing of smaller voids embedded in large-scale collapsing overdensities responsible for the decrease in void counts at later times, as can be seen in Table~\ref{tab:voidcounts}.

Overall, these numbers provide sufficient statistics to perform an analysis of the void properties in the context of the halo mass function in voids, and as such, we will first investigate the properties of the voids in general before diving into the details of the VHMFs. 
  
\subsection{Void Property Distributions}\label{sec:voidprop}
\begin{figure*}
   \centering
   \includegraphics[width=0.9\textwidth]{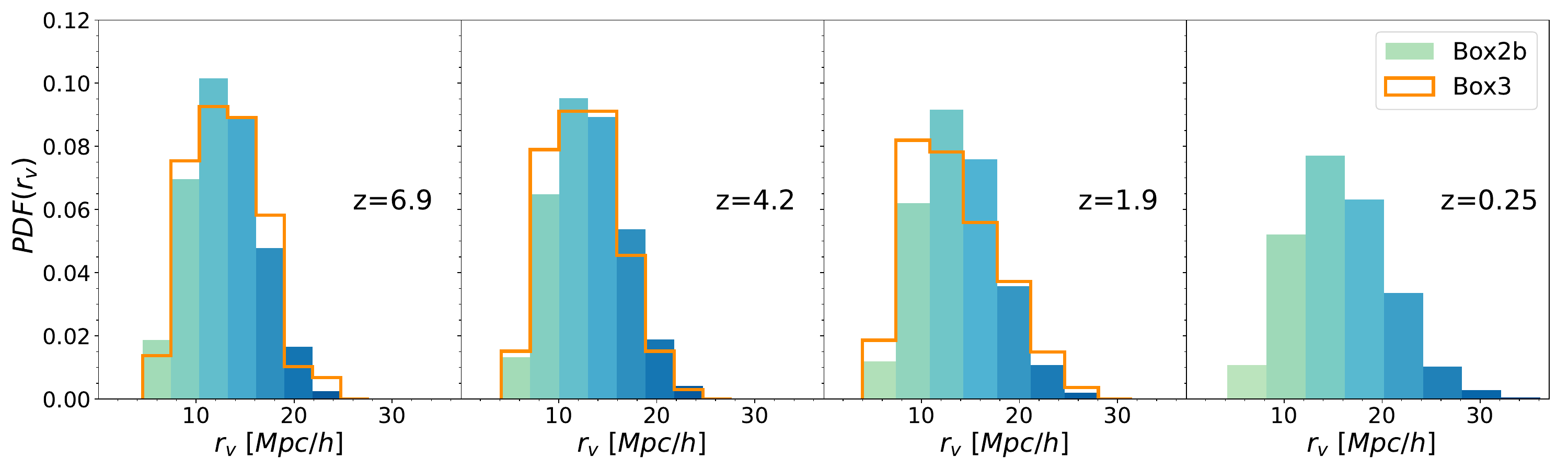}
    \includegraphics[width=0.9\textwidth]{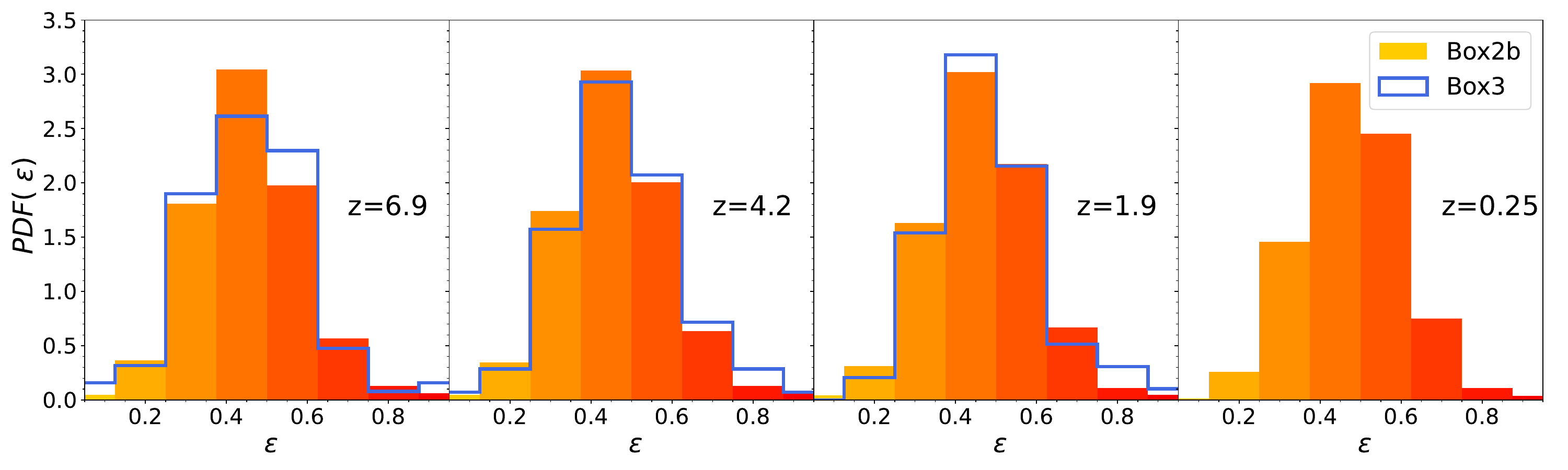}
    \includegraphics[width=0.89\textwidth]{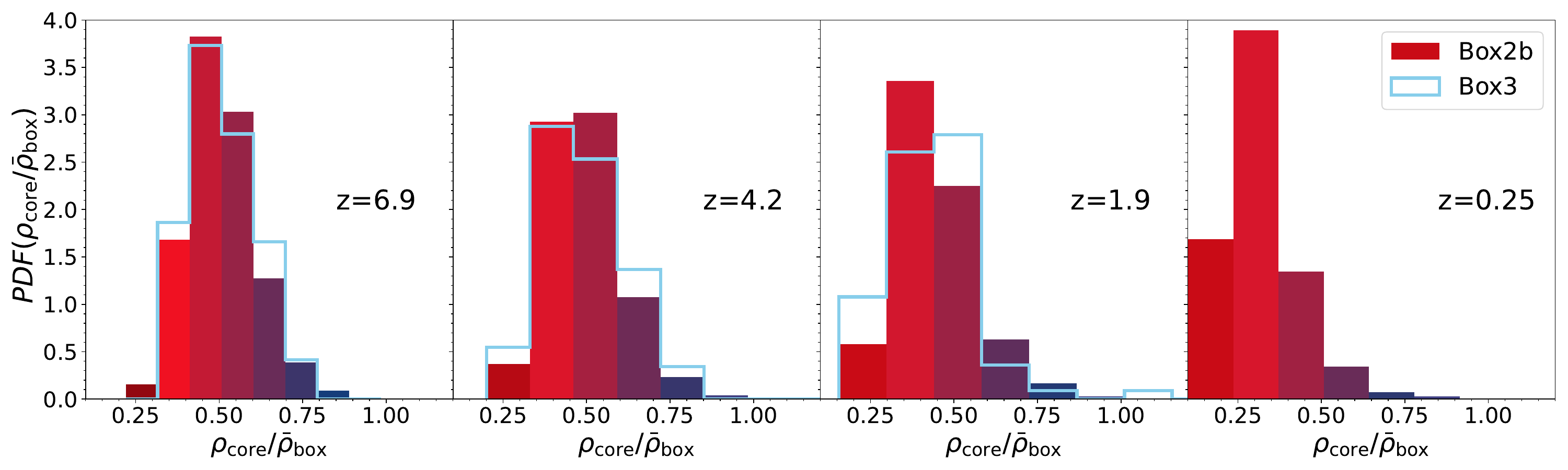}
     \caption{Histograms of the void properties (void-size, ellipticity, and core density from top to bottom) at four different redshifts of $z=6.9$, $z=4.2$, $z=1.9$, and $z=0.25$, from left to right. The colored histograms show the result for \textit{Box2b}, while the resulting distributions from \textit{Box3} are shown as orange or blue lines for all available redshifts. Note that the color gradient of the \textit{Box2b} histograms has no physical meaning; however, it will be used to highlight different void-sizes later on. The last row shows the core density $ \rho_{core}/ \bar{\rho}_{box}$ distributions, with $\rho_{core}$ describing the density in the most underdense Voronoi-cell and  $ \bar{\rho}_{box}$ is the mean comoving density of the simulation volume, which stays constant with redshift.}

   \label{fig:prop_hist}%
\end{figure*}

\begin{figure*}
   \centering
   \includegraphics[width=0.33\textwidth]{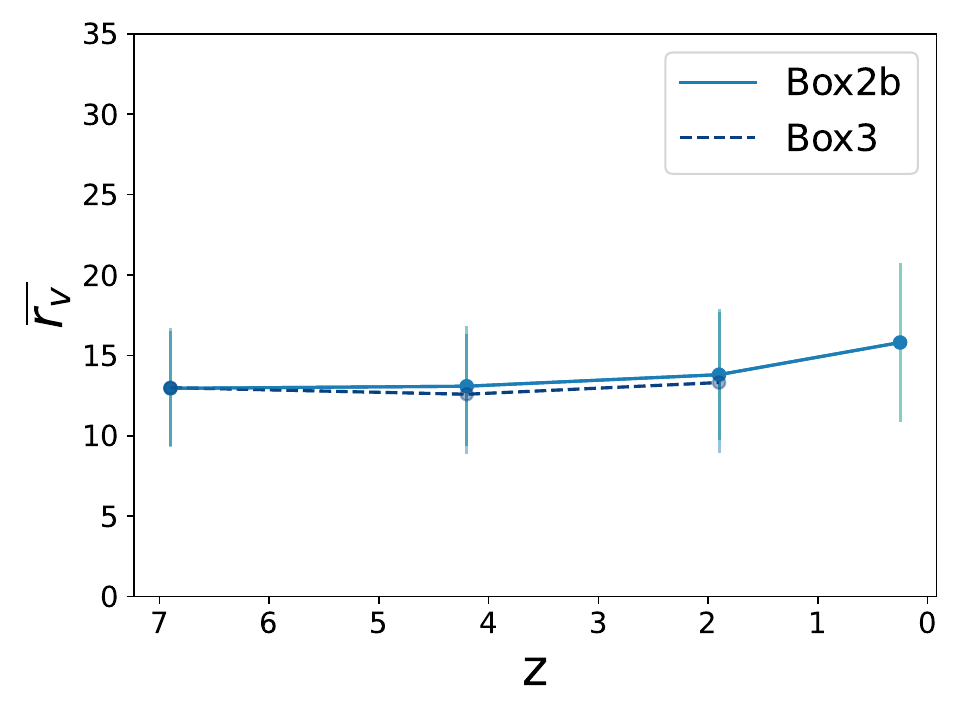}
   \includegraphics[width=0.33\textwidth]{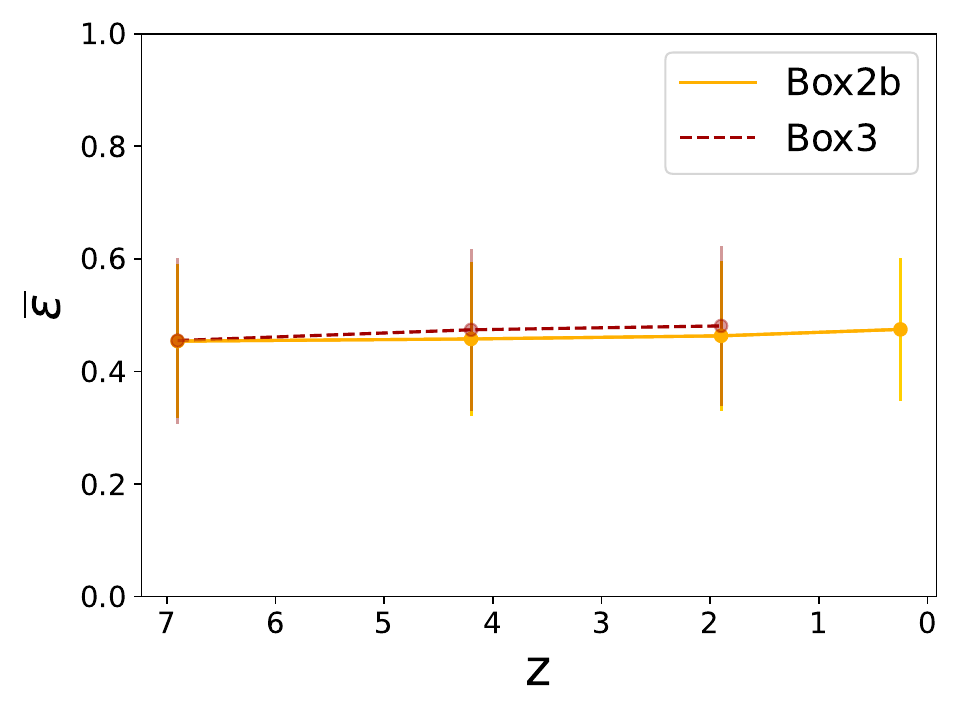}
   \includegraphics[width=0.33\textwidth]{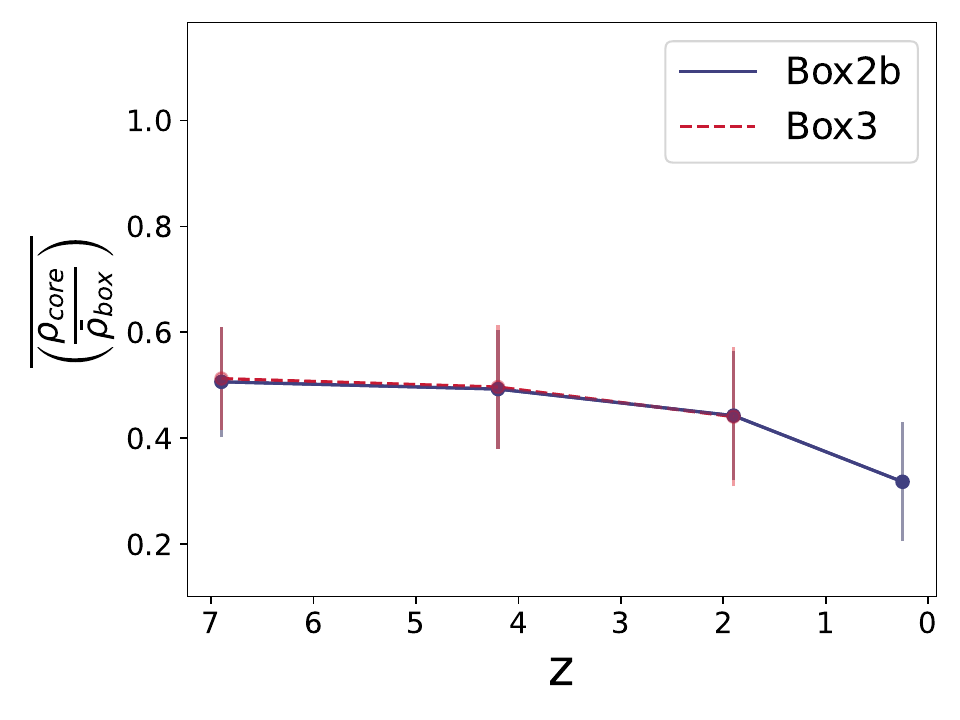}
   \caption{Time evolution of the mean void radius with the standard deviation in light blue on the left. The time evolution of the mean ellipticity with the standard deviation in light orange in the middle panel and the core density in purple on the right, both calculated in \textit{Box2b} as well as \textit{Box3} (dashed line). While the mean ellipticity and radius remain stable throughout the high redshift universe, with a slight upward trend towards redshift $z=0$ for the void radius, the core density shows an inverse trend and undergoes a more pronounced evolution earlier on. }
   \label{Fig:zevo}
\end{figure*}
As discussed in the previous section, the number of voids decreases with time as voids evolve, when large voids merge, and small voids collapse (if surrounded by an overdensity) \citep{sheth_hierarchy_2004}. But does this also imply that the voids themselves are growing in size with redshift?  We investigate this question in Fig.~\ref{fig:prop_hist}, where the void size distributions for both simulation volumes are shown at four different redshifts from $z\approx 7$ to $z=0.25$ in the first row. As one can see, the maximum void sizes in the simulation increase toward lower redshifts, reaching radii larger than $30~\textrm{kpc}$ at $z=0.25$. The range of the void sizes at a redshift of $z=6.9$ spans from $r_v = 4.56$ to $r_v = 27.66$, while at a redshift of $z = 0.25$ the radii lie within an interval of $r_v = 4.21$ to $r_v = 36.15$. This indicates that the voids are indeed growing in size over time, visible in \textit{Box2b}, colored in blue, as well as in \textit{Box3}, depicted by the orange lines in the upper row of Fig.~\ref{fig:prop_hist}. This can also be seen in the left panel of Fig. ~\ref{Fig:zevo}, where the mean void radius is shifted toward a higher value for a redshift of $z=0.25$, while remaining rather stable for earlier cosmic times of redshift $z \approx 7$ to $z \approx 2$. These findings also correlate closely with Fig. ~\ref{fig:boxview}, where one can see the structure of the cosmic web becoming more pronounced with time.

Voids are not only characterized by their size, but also by their shape. As VIDE is a shape-free method, assuming no a priori void shape (i.e., spherical or ellipsoidal voids), the void population can be quite diverse. Here, we quantify the shape of a void by its ellipticity as introduced in section~\ref{sec:vide}.

We investigate the evolution of the ellipticity distribution in time in the second row of Fig.~\ref{fig:prop_hist}. Contrary to the evolution of the void radius, the ellipticity shows a less pronounced evolution, remaining quite stable around $\bar{\epsilon} \approx 0.45$. There is a slight hint of a higher mean ellipticity towards redshift $z=0.25$, which might stem from the increasingly defined cosmic web further constraining the shape of the void.

Possibly the most interesting void property is the core density, as it gives a direct estimate of the local underdensities in the core of the void compared to the mean density in the total simulation volume. Therefore, the property quantifies the most underdense region of the void, usually located around the void center, normalized by the mean co-moving density of the whole simulation volume. 
The last row of Fig.~\ref{fig:prop_hist} shows the evolution of the core density with the mean depicted in the right panel of Fig.~\ref{Fig:zevo}. We see a continuous decrease in core density, again, most pronounced from $z=2$ to $z=0$. The evolution of this parameter is the strongest, indicating that voids become increasingly "empty" with cosmic time.\\

Comparing the results of the larger, lower-resolution \textit{Box2b} with the smaller, higher-resolution \textit{Box3}, we see good agreement in the property evolution trends, indicating no significant deviations due to resolution effects in void size, shape, and density-contrast. This is further justification for moving forward with the larger, yet less resolved \textit{Box2b} for the study of the effect that void properties have on the VHMFs.

Any of these properties of the voids, as well as their evolution in time, could reasonably impact the evolution of the galactic halos within them. In particular, the ellipticity and steepness of the density profile trace the local tidal fields \citep{park2007a} within these regions and consequently the environment in which the halos grow up.
Ultimately, this leads to the question of how these trends affect the Void-Halo-Mass-Function. We will further investigate this in Section ~\ref{sec:VHMF_voidprop}.

\subsection{The Void Halo Mass Function}\label{sec:VHMF}


\begin{figure*}
   \centering
   \includegraphics[width=0.7\textwidth]{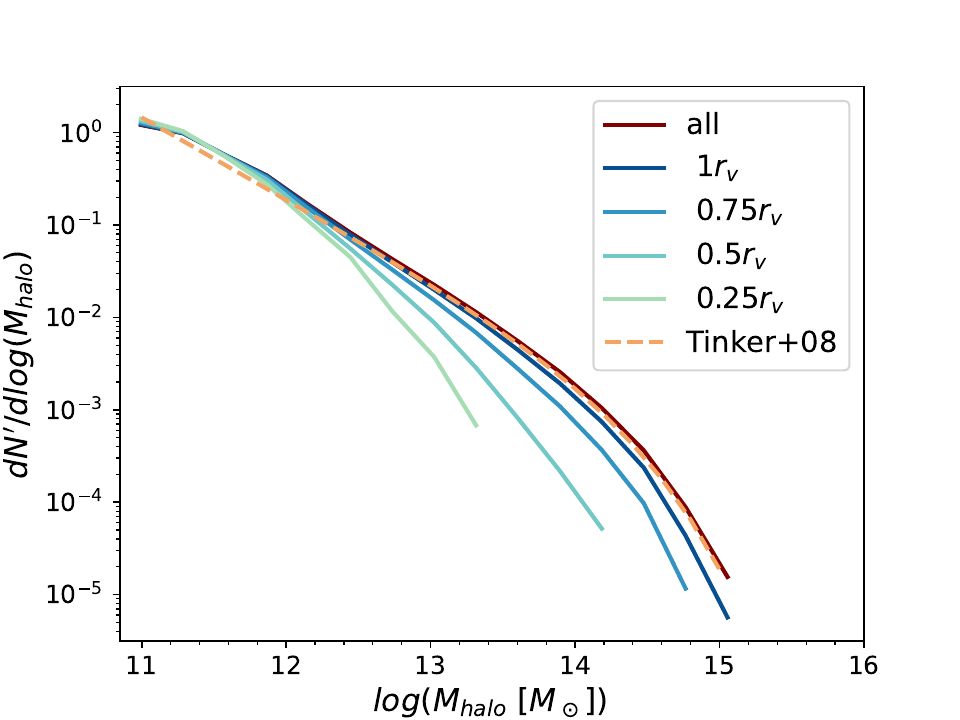}
   \caption{ Halo mass function within various void-radii at a redshift of $z=0.25$ in \textit{Box2b}. The lightest shade of turquoise includes all halos that lie within $0.25$ of the radius of the respective void $r_v$. With increasing darkness of the color, more halos further from the void center are included. $1r_v$ takes all halos that lie within one void radius into account when calculating the void halo mass function. The maroon line labeled \textit{"all"} is the standard halo mass function obtained from all halos in \textit{Box2b}. }
              \label{fig:VHMF}%
\end{figure*}

In order to assess whether and how the halo populations differ inside voids, we use the Halo-Mass-Function $\rm{dN'/dlog(M_{halo})}$ as a main tool. Here, $\rm{N'}$ is the relative cumulative frequency of halos counted within a certain logarithmic mass range $\rm{log(M_{halo})}$ divided by the total number of halos. Each VHMF is then calculated by considering only halos that lie within their respective voids rather than by including the complete halo catalog of each simulation volume. We calculate a VHMF for different fractions of spherical void radii $r_v$, namely, volumes of $0.25r_v, 0.5r_v, 0.75r_v$ and $1r_v$ for each void, introducing a set of four VHMFs. This allows us to gain a deeper understanding of how the HMF varies with the density of the environment. 
Fig.~\ref{fig:VHMF} shows the set of VHMFs at a redshift of $z=0.25$, ranging from $0.25r_v$ in pale green to $1r_v$ in successively darker shades of blue. The general HMF is shown in maroon red, labeled "all", and the dashed orange line depicts the HMF derived from the Extended-Press-Schechter formalism \citep{bond_excursion_1991} using the expanded HMF model with fitting parameters according to \citet{tinker:2008}. Consistent with expectations, the general mass function is in close agreement with the theoretical HMF model.  For the different VHMFs, one can see a split-up in the high-mass end in Fig.~\ref{fig:VHMF}.
This indicates that the most underdense parts of the void, only one fourth of the void radius from the center, restrict the halo mass at $\rm{log(M_{halo})} = 13.315 $. 
As the outer parts of voids generally contain more mass than the inner parts, this maximum halo mass increases to $\rm{log(M_{halo})} =  15.055$ at $1r_v$, most closely matching the general HMF while still showing a comparably lower number of high-mass halos.

This implies that at the edges of voids, where the total density of the region approaches the mean density of the universe via an overdense compensation wall, it is again possible for galaxy cluster mass halos to form. However, the relative frequency of these halos has not yet approached the general HMF because the overall volume is still dominated by the less massive void halos.

Summarizing the findings shown in Fig.~\ref{fig:VHMF}, it is apparent that cosmic voids, which are generally lower in density, host less massive galaxies.
The deeper inside the void, the more restrictive the environment, and the more noticeably favored are low mass halos.

Apart from the split-up of the VHMFs, another interesting property is the slight overabundance of low-mass halos within Voids, as reflected in the change in overall slope. Fig.~\ref{fig:resolution} shows this more clearly, as we show the same figure for the smaller but higher resolution simulation, as the higher resolution enables us to see the trend to lower masses. Paired with a normalization matching that from \textit{Box2b} emphasizes this effect (for more details see Appendix ~\ref{app_res}). 
This shows in the low-mass-end in Fig.~\ref{fig:VHMF} around $\rm{log(M_{halo})} = 11$, where the halo-count is the highest for $0.25r_v$ with a slight decrease with increasing radius from the center of the void.
In conclusion, Fig.~\ref{fig:VHMF} shows that at a redshift of $z=0.25$ the VHMFs diverge from the general HMF, where the most underdense part of the void shows the strongest mass cut-off and simultaneously exhibits the highest relative overabundance of low-mass halos as a result of the steeper slope. The mass threshold is gradually allowing more massive halos to form as the radius $r_v$ from the void center increases.  
   

\begin{figure*}
   \centering
   \includegraphics[width=0.9\textwidth]{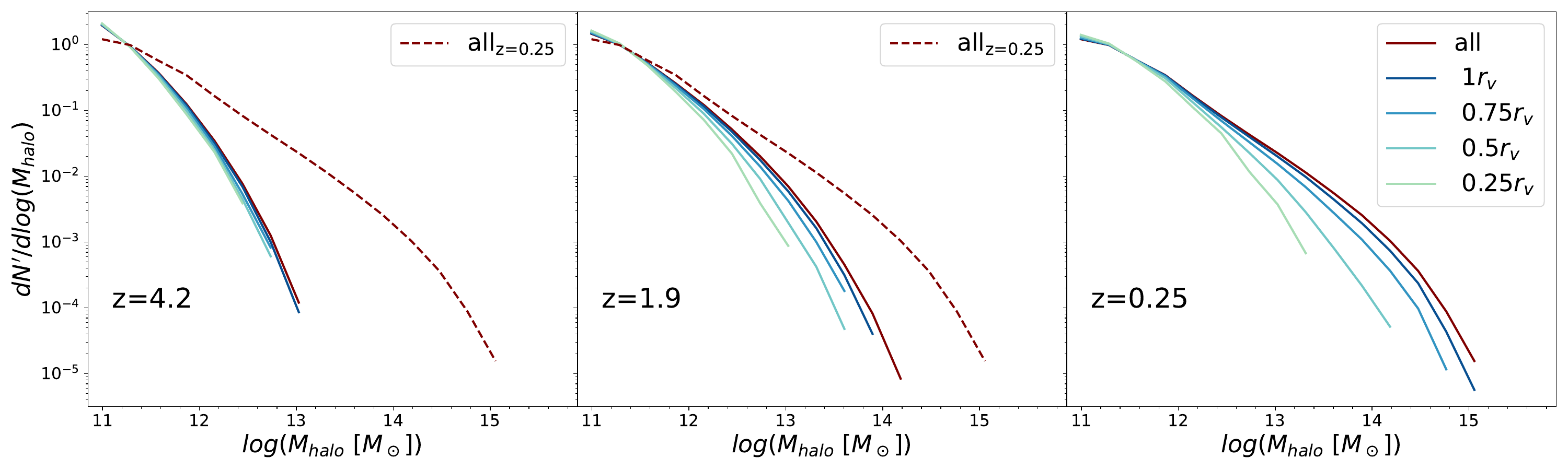}
   \caption{Evolution of the VHMFs over time in \textit{Box2b} with decreasing redshift of $z=4.2$, $z=1.9$ and $z=0.25$ from left to right. In the left and middle panels, the maroon dashed line shows the HMF at redshift $z=0.25$, allowing us to easily see the evolution of the split-up of the VHMF with advancing redshift. At a redshift of $z=4.2$, there is no significant divergence of the different VHMFs visible, correlating with the still rather homogeneous universe. As time evolves, the splitting of the VHMs becomes more apparent as the density contrast is amplified with voids becoming more underdense and already dense regions accumulating more mass.  }
              \label{fig:vHMF_evo}%
\end{figure*}

Since we find a divergence among the VHMFs, it follows naturally to look into the time evolution of the split-up. In accordance with Section ~\ref{sec:voidprop} we investigate redshifts of $z=6.9$, $z=4.2$, $z=1.9$ and $z=0.25$. However, at a redshift of $z=6.9$, there are not enough halos present within the resolvable mass range to create viable statistics. Therefore, Fig.~\ref{fig:vHMF_evo} shows the VHMFs at redshifts $z=4.2$, $z=1.9$, and $z=0.25$ from left to right, respectively, following the same color scheme as Fig.~\ref{fig:VHMF}, marking the different spherical shells from the void center. The dashed line represents the general HMF at redshift $z=0.25$, labeled "all" in the right panel for reference. In the left panel, the VHMFs do not exhibit a significant split-up, although there is a faint hint that the VHMFs of $0.25r_v$ and $0.5r_v$ already host less massive galaxies than the outer parts of the void. The split-up starts to appear in the middle panel at $z=1.9$ and becomes more pronounced at $z=0.25$, as previously described. With an increasingly apparent split-up of the VHMFs from left to right, the overabundance of low-mass halos also becomes slightly more pronounced as not only the walls and filaments are evolving to become increasingly apparent, but also the density contrast inside voids intensifies as shown in the lowest rows of Fig.~\ref{fig:prop_hist}. This is in accordance with Fig.~\ref{fig:boxview}, where the matter is distributed rather homogeneously at a redshift of $z=6.9$, and the density contrast becomes increasingly prominent as mass starts to agglomerate along the cosmic web. In conclusion, Fig.~\ref{fig:vHMF_evo} shows that as the cosmic web starts to form, the VHMFs evolve, showing a split-up and a slight overabundance of low-mass halos in voids.

\subsection{Dependence of the VHMF on Void Properties} \label{sec:VHMF_voidprop}

\begin{figure*}
   \centering
   \includegraphics[width=.9\textwidth]{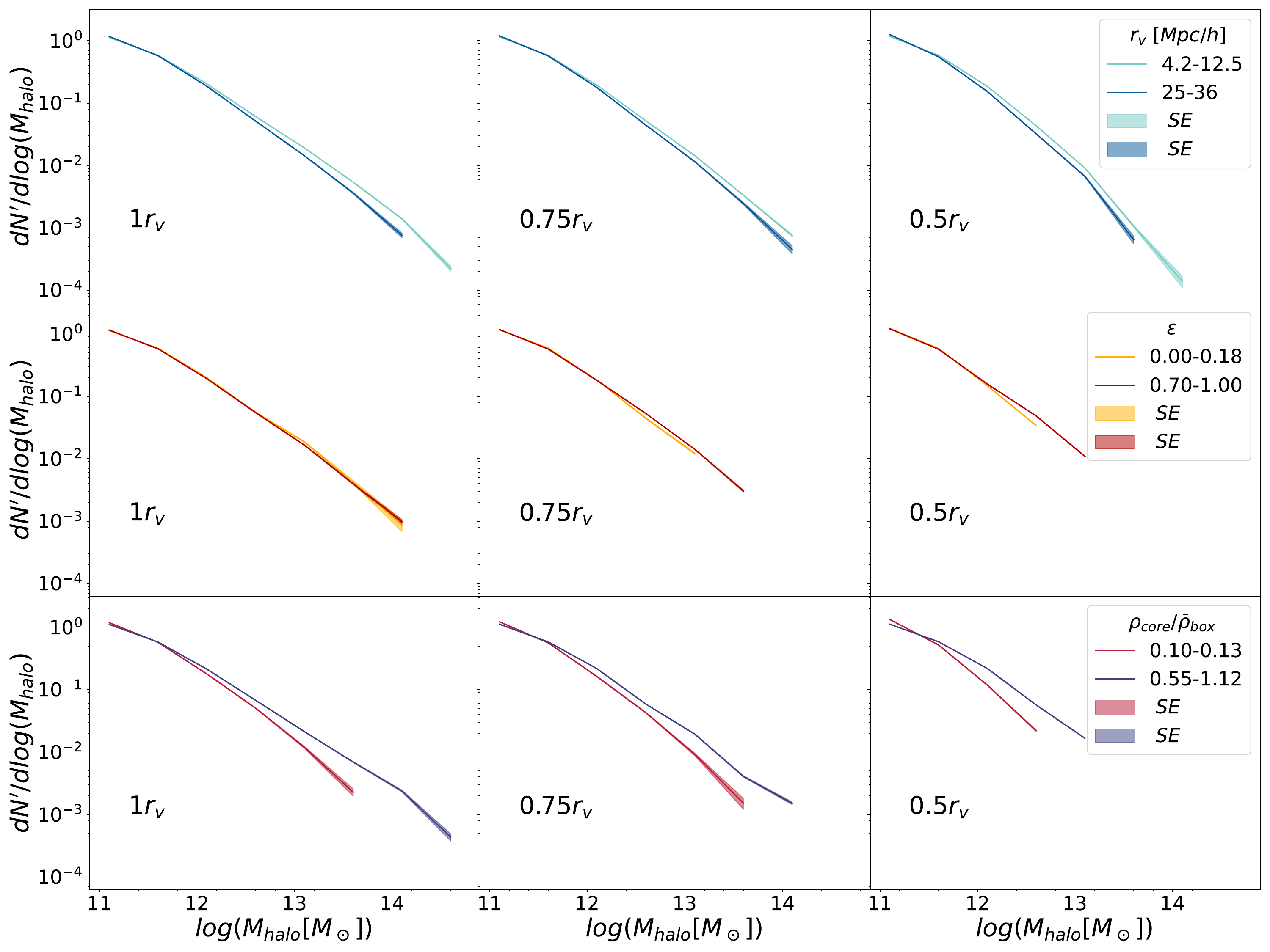}
   \caption{Impact of void properties on the VHMFs. $1r_v$, $0.75r_v$, and $0.5r_v$ VHMF are shown from left to right, each calculated from two different halo-catalogs depending on the three void properties: radius, ellipticity, and density-contrast from top to bottom. The standard error is indicated as a shaded area around the lines depicting a small statistical sampling error, though this is not clearly visible for every VHMF. }
              \label{fig:splitmaxmin}%
\end{figure*}
Returning to the void properties discussed in Section ~\ref{sec:voidprop}, we now investigate whether they impact the VHMFs. To this end, we subdivided all VHMFs according to the size, ellipticity, or density contrast of the voids in \textit{Box2b} at redshift $z=0.25$.

To show the maximum possible effect void properties can have on the VHMFs, we directly compare the 95th and 5th percentiles of voids in each property.
We do this for the VHMFs that are depicted in Fig. ~\ref{fig:VHMF} except for the $0.25r_v$ VHMF since it does not have a large enough halo-catalog. However, a three-equal-sized bin split-up is shown in Appendix Fig.~\ref{fig:split} where the $0.25r_v$ VHMF is included.
Fig. ~\ref{fig:splitmaxmin} shows the VHMFs $1r_v$, $0.75r_v$, and $0.5r_v$ from left to right with void-size, ellipticity, and density-contrast from top to bottom. 
\subsubsection{Void-Size}
The first row of Fig.~\ref{fig:splitmaxmin} shows the VHMFs calculated using a halo catalog that includes all halos that reside in voids with sizes of $r_v = 4.2-12.5 $ in light blue and $r_v = 25-36$ in dark blue with their respective standard error. Generally, they converge with a hint that larger voids host less massive halos. One reason for this trend could be that larger voids arise from steeper initial density fluctuations and hence evolve to be a less dense environment, which, as seen in Sec. ~\ref{sec:VHMF}, affects the VHMFs. \cite{schuster_nonlinear_2023} showed a correlation between the void radius and the core density, indicating that larger voids are typically less dense at the core. 

\subsubsection{Ellipticity}
The second row of Fig. ~\ref{fig:splitmaxmin} displays two VHMFs each, now calculated from a halo catalog that includes halos residing in voids with ellipticities of $\epsilon = 0.00-0.18$ in yellow and $\epsilon = 0.7-1.0$ in red. 

\begin{figure*}
   \centering
   \includegraphics[width=.9\textwidth]{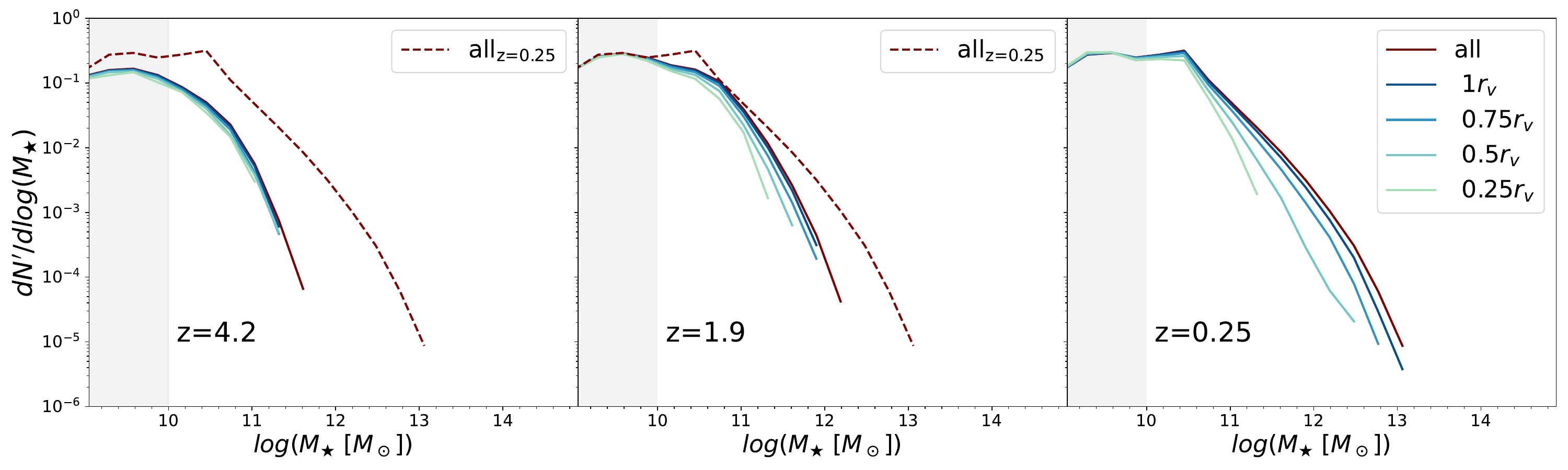}
   \caption{Evolution of the stellar mass function from redshift $z=4.2$ to $z=0.25$ from left to right. The dashed maroon lines at redshifts $z=4.2$ and $z=1.9$ serve as reference lines for the solid dark blue line in the rightmost panel, which shows the stellar mass function at a redshift of $z=0.25$.}
              \label{fig:smfevo}%
\end{figure*}
The two VHMFs are in strong agreement, indicating that the ellipticity of the void has no effect on the VHMFs. This finding reinforces that using spherical radii for filtering the void halos is a good first-order approximation, since otherwise the highly elliptical voids would be strongly contaminated in their VHMFs. It should be noted that the assumption still results in a slight oversampling of high-mass halos in the outskirts of the voids since it includes some halos that are part of the cosmic web. Nevertheless, the qualitative findings are still valid, and a slightly steeper split up between the general HMF and the $1r_v$ VHMF (Fig.~\ref{fig:VHMF}) would be expected if the ellipticity of the void were taken into account more accurately. 
\subsubsection{Core Density}
Following the same method as previously described, the third row of Fig. ~\ref{fig:splitmaxmin} shows two VHMFs per panel with a density-contrast of $\rho_{core}/\bar{\rho}_{box} = 0.10-0.13$ (raspberry red) and $\rho_{core}/\bar{\rho}_{box} = 0.55-1.12$ (blue violet). A lower core density implies a more underdense region compared to the mean density of the simulation volume. Among all void properties, the core density of individual voids most strongly affects the hosted galaxies. Each VHMF (from $0.5r_v$ to $1r_v$) shows a split-up between the two core-density-selected samples, seemingly independent of the distance to the void center. Voids with a higher core density host more massive halos. This is to be expected, as voids with low core densities arise from deeper initial density fluctuations and hence grow to become more pronounced compared to high core density voids. On the other hand, this result shows that the VHMFs are sensitive not only to local densities but also to the void's core density.

\section{Impact of Voids on Baryonic Processes}

We have established that cosmic voids manifest as a limiting factor for the halo mass located in a void compared to field halos, as well as also inducing a threshold on the stellar mass. In the following, we will look at the stellar-mass-halo-mass relation as well as the main sequence of star-forming galaxies to answer the question of how voids impact baryonic physics on galactic scales and, if so, what insight we can gain from it. 

\subsection{Void-Stellar-Mass-Functions}
An important scaling relation for the study of galaxy formation is the distribution of stellar masses among galaxies, known as the stellar mass function (SMF). It ties the purely gravitational mass of a halo to the stellar mass impacted by the baryonic physics at play in the galaxies themselves. Therefore, it is the ideal tool for studying whether the suppression of halo growth observed in the previous sections also affects the dynamics at smaller scales. Additionally, it is much more observationally accessible than the HMF and has been studied by various observational campaigns \citep[e.g.][]{muzzin:2013,weaver:2023}. As shown by \cite{dolag2025}, for the resolution used in this work, the SMF agrees well with observations. We therefore investigate the behavior of this relation in the Magneticum voids.
Fig. ~\ref{fig:smfevo} shows the evolution of the stellar mass function in \textit{Box2b} from redshift $z=4.2$ to $z= 0.25$ from left to right, analogous to Fig.~\ref{fig:vHMF_evo}. Any values in the gray-shaded area are beyond the resolution limit of the simulation volume. The SMF (labeled "all" in maroon-red) and the void stellar mass functions (VSMF) are colored as previously, from pale green to dark blue. The dashed line in the left and middle panels of Fig.~\ref{fig:smfevo} represents the SMF at redshift $z=0.25$ labeled "all" in the right panel.
Similar to the HMF in Section ~\ref{sec:VHMF}, we see an evolving split-up of the VSMFs with declining redshift for galaxies with stellar masses larger than $10^{11} M_{\odot}$, while for smaller galaxies we do not see differences among the VSMFs. The overabundance at the low-mass end is less prominent than for the VHMFs. and is completely missing for small galaxies. This opens the question if the stellar-mass-halo-mass relation could be different in voids.

\subsection{Stellar Mass-Halo Mass Relation in Voids}
\begin{figure*}
    \centering
    \includegraphics[width=.9\textwidth]{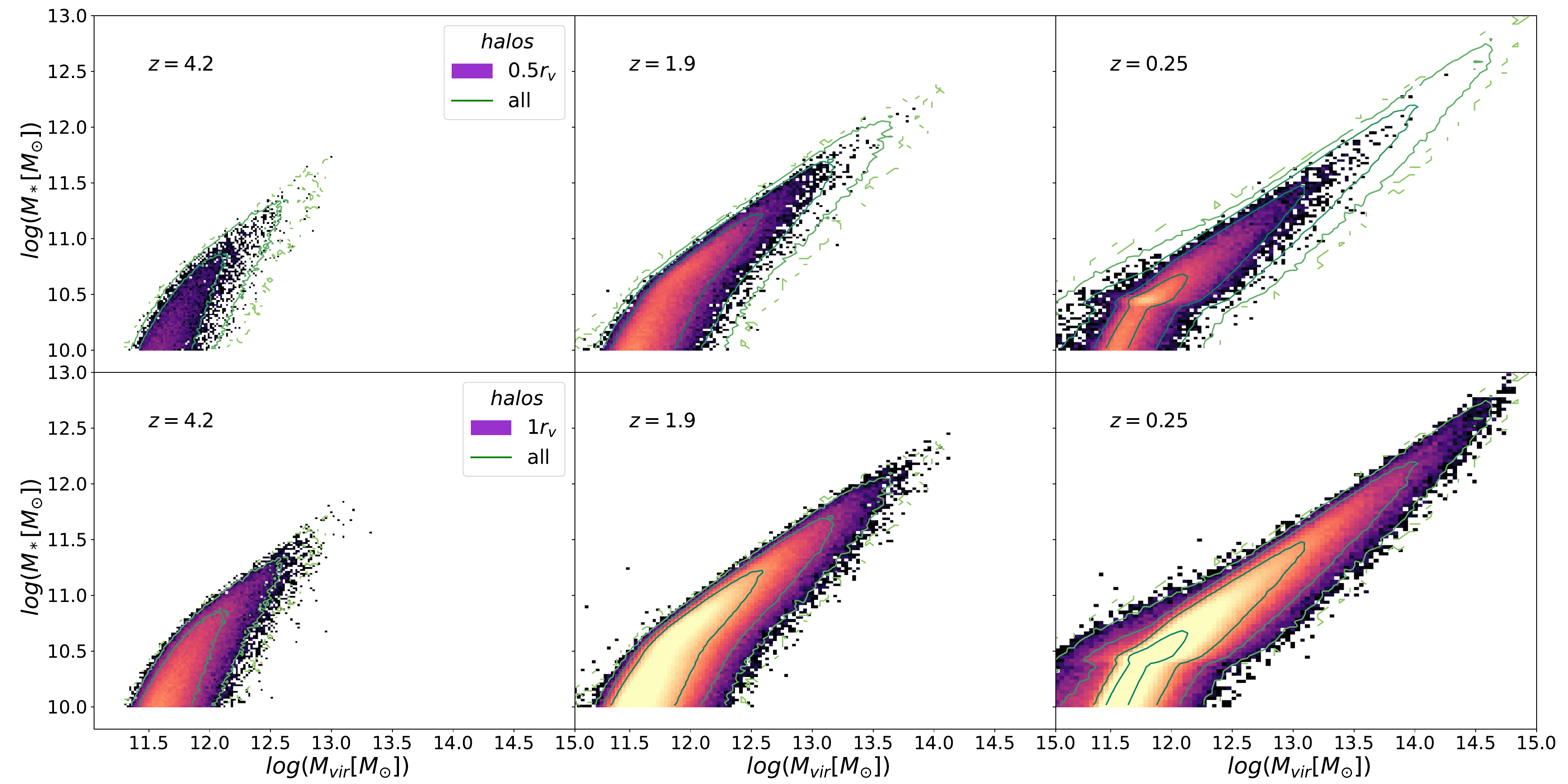}
    \caption{Stellar-mass-halo-mass relation throughout redshifts $z=4.2, 1.9, 0.25$ from left to right, where the green contours always represent the full galaxy catalog. The top panel shows galaxies within $0.5r_v$ in purple to yellow, while in the bottom panel, the colors represent the full void galaxy catalog. For both, the void galaxies follow the contours and show only a limited halo and stellar mass. }
    \label{fig:hm-sm}
\end{figure*}

Figure ~\ref{fig:hm-sm} shows the stellar mass-halo mass relation at redshifts $z=4.2, 1.9, 0.25$ from left to right. The top panel depicts the entire halo population in \textit{Box2b} as green contours, and all halos within $0.5r_v$ are shown by the heatmap ranging from dark purple to yellow. The bright yellow color represents a high halo count per pixel ($\geq 1000$), whereas the dark purple shows the scatter of the relation with only one halo per pixel. For the top panel, the halos within $0.5r_v$ of their respective voids show lower stellar as well as virial mass through all redshifts, while the shape agrees with the contour of all halos in the simulation volume.
The same trend is also visible in the bottom panel, where the green contours again show the entire halo population, and the heatmap represents halos residing within $1r_v$, thereby showing the whole void-halo population. This suggests that on a statistical scale, there are no systematic differences between void halos and the general population. However, as shown in the previous sections, the inner parts of the void can be interpreted as a mass-restricting component for halos located within it.     

\subsection{The Star-Forming Main Sequence in Voids}

\begin{figure*}
    \centering
    \includegraphics[width=.9\textwidth]{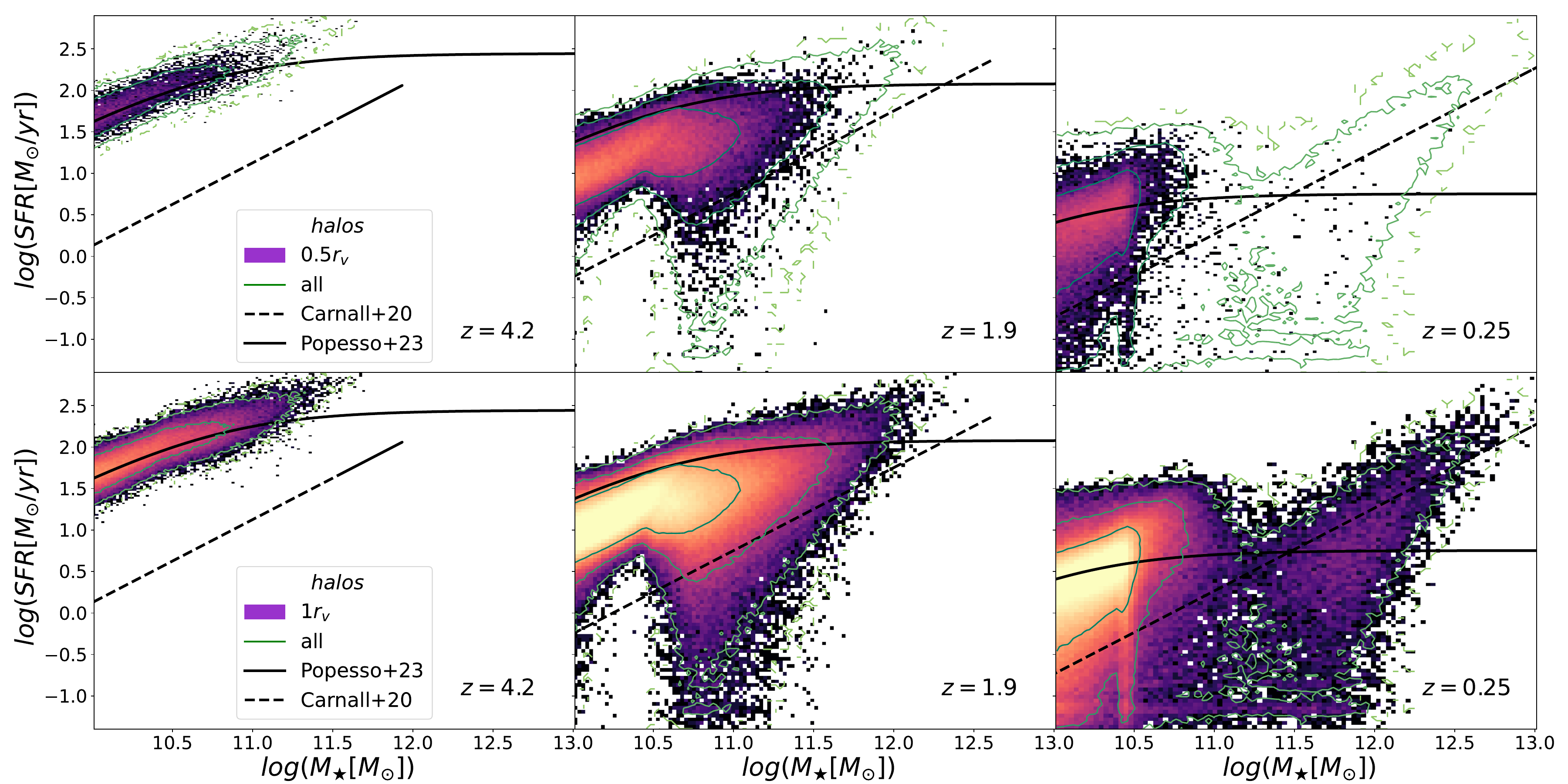}
    \caption{Star-forming main sequence for the full halo catalog shown by the green contours at redshifts $z= 4.2, 1.9, 0.25$ from left to right. The black line shows the criterion for quenched star formation below which galaxies are considered to be quiescent as proposed by \cite{Carnall_2020}. Analogous to Fig.~\ref{fig:hm-sm}, the top panel shows void galaxies within $0.5 r_v $ and all void galaxies are shown in the bottom panel in purple to yellow.}
    \label{fig:sfr}
\end{figure*}

Figure ~\ref{fig:sfr} shows the star-forming main sequence again at redshifts $z=4.2, 1.9, 0.25$ from left to right. Analogous to Figure~\ref{fig:hm-sm}, the top and bottom panels show heatmaps of void halos within $0.5r_v$ and $1r_v$, respectively. The green contours show the stellar mass - star formation rate plane of the whole halo population within \textit{Box2b}. As a reference, the dashed black line depicts the criterion for star-forming galaxies given by $sSFR \geq 0.2/t_H$ as used by \cite{Carnall_2020}, scaling with the Hubble time $t_H$. The simulations are in good agreement with \cite{Popesso_2023}, shown in the solid black line. The colors of the heatmap depicting the void halos are the same as in Figure~\ref{fig:hm-sm}.
Until now, there has been no clear consensus about the star formation in void galaxies. While \cite{rojas:2004} find a higher specific star formation rate for galaxies of a given mass and  \cite{rodriguezMedrano:2024} predict a higher absolute star formation rate, \cite{kreckel:2012} find no significant difference in the star formation between void galaxies and the general population.
This work agrees with the latter, as in \textit{Box2b} of the Magneticum simulations, the contours are in good agreement with the void halos. It should be noted that, by taking a statistical approach to the matter, we are unable to resolve low-mass halos. As such, there is a possibility that low-mass halos behave differently. However, for halos with stellar masses of $M_* = 10^{10}$ or higher, there is no difference between void and field halos in terms of their SFR or stellar-mass-halo-mass relation. In conclusion, voids seem to have a mass limiting effect only without affecting common scaling relations for galaxies with stellar masses of $M_* > 10^{10}M_{\odot}$. However, at a redshift of $z=0.25$ within $0.5$ of the void radii, star-forming galaxies are found with masses up to $\approx 10^{10.5} M_{\odot}$ whereas Fig.~\ref{fig:hm-sm} still shows stellar masses of $\approx 10^{11.7} M_{\odot}$ at the same redshift and void radius. Fig.~\ref{fig:quenched_sf_vHMF} shows more clearly that inside voids, the quenched fraction is higher than for the general population. This effect is most visible for the $0.25r_v$ VHMF as it enforces a strict cutoff for star-forming galaxies at a mass of $\approx 10^{10.4}$, while for the other VHMFs, this is a trend rather than a criterion. This highlights that the innermost regions of voids already limit not only mass but also star-forming activity in the local galaxies. It is to be expected that, as time progresses, the strict cutoff will also appear for the $0.5r_v$ VHMF as the on-site gas will be used up, and there is no mechanism to replenish those regions as matter couples to the large-scale structure and leaves low-density regions devoid of new baryonic matter.

\section{Discussion}
The importance of the local environment on the formation of halos has long been studied. Implications such as higher star formation and more blue, rather late-type spiral galaxies to be found in voids are still a topic of debate. (e.g., \cite{rojas:2004},\cite{rojas:2005} \cite{kreckel:2012},\cite{rodriguezMedrano:2024}). 
This work shows that cosmic voids act as a mass-limiting factor at a redshift of $z=0.25$, which changes the slope of the HMF depending on the radial distance to the void center. However, this trend is not present across all redshifts as the split-up only starts to evolve around $z \gtrsim 2$. This trend is closely tied to the assembly of the cosmic web and to the steepening density contrast within the simulation volume as redshift decreases. One could, therefore, quantify the level of homogeneity on scales ranging from $\sim 4$ to $\sim 30$ Mpc/h. At a redshift of $z=4.2$, the universe presents inhomogeneities only on scales smaller than $4$ Mpc/h. Note that this is dependent on the tracer density that is used to identify voids as shown by \mbox{\cite{sutter_sparse_2014}}.\\  

Void galaxies also gained importance in the field of precision cosmology and have been used to further constrain $\Omega_m$ and $\sigma_8$ (\cite{contarini_2023}). 
In this work, we highlight the impact of global void-properties -- void-radius, ellipticity, and core density -- on the hosted halo population. In the framework of a set of four spherical VHMFs, we find a sensitivity of the HMF to underdensities. This suggests that VHMFs could be used to trace the current state of the cosmic web, which, depending on redshift (or cosmology), could also allow for a more precise tracer of cosmological parameters. 
Since this also applies to the VSMFs, it could pose a useful tool in assessing large tracer-dense galaxy surveys. The point in time at which the split-up starts to occur can also be used to obtain information about $\Omega_m$ and $\Omega_{\Lambda}$ equality. However, the tracer density is currently not sufficient in galaxy surveys at high redshifts.   

Crucially, the impact of the core density of the probed voids should not be underestimated, and to obtain physical insights, the calibration of the derived HMFs needs to be handled with caution. 
\cite{pujol_what_2017} found that density is also the driving bias in halo occupation distribution models, contrary to the assumption that mass is the determining factor. Since we find that voids with a high core density show a shallower slope in their VHMFs compared to voids with a low core density, this indicates that, depending on the local environment of the void, the halo mass threshold varies to a significant extent. So not only is the HMF sensitive to the environment within voids, but each VHMF is additionally dependent on the core density of the traced void population. 

A more in-depth study of the impact of voids on the baryonic processes in this work shows that the stellar-mass-halo-mass relation for the void halos within $0.5 r_v$ as well as the $1r_v$ falls within the contours of all halos (where the void halos only make up $27 \%$ of the total number for the $1r_v$ case). This shows that, in a statistical sense, there are no different physical processes involved in void halos compared to the general population. It should be noted that the resolution of the dark matter and gas particles could play a role, and smaller halos may behave differently. An impact cosmic voids do have on the star formation rate is that void galaxies at higher stellar masses tend to stop their star formation at $z=0.25$ compared to the general galaxy population which are able to prolong their star forming activities also for galaxies with stellar masses larger than $\rm{log(M_* [M_{\odot}])}\approx 10.7$ along the whole mass range. In that sense, a possible explanation could be that, as matter evacuates from voids over time, gas also becomes scarcer earlier on. However, various processes determine a galaxy's potential to form stars, and hence it is not clear whether this simple picture can explain this result.

\section{Conclusions}\label{sec:conclude}

We used \textit{Box2b} from the Magneticum simulations to investigate the evolution of void properties and their effect on the HMF, as well as the impact of voids as a probe of underdense environments on the HMF. The main findings can be summarized as follows:

\begin{itemize}
  \item The global properties of voids evolve over time, with the radius of the voids showing the most development, since voids grow through merging and can be squished by surrounding overdensities if they are small. This results in a decrease in the total void number inside the simulation volume and a shift in the mean radius towards larger voids at lower redshifts. The ellipticity shows little evolution across redshift. However, the mean core density shifts toward lower values. This means that voids become more underdense with time as the comic web assembles. Note that the evolution of the mean is not linear, but the shift occurs between redshifts of $z=2$ and $z=0$. 
  
  \item At a redshift of $z=0.25$, the VHMFs show a split-up in the high-mass-end as well as a slight overabundance of low-mass halos. The closer to the void-center, the higher the overabundance of low-mass halos, and the fewer high-mass halos can be found. 
  The VHMF closest to the void center shows the fewest high-mass halos. 
  This shows that the distribution of galaxies is closely tied to the environment. 

  \item It follows that the split-up between the VHMFs is evolving in time, closely tied to the assembly of the cosmic web. At a redshift of $z= 4.2$, the matter distribution in the universe is rather homogeneous and therefore voids have no impact on the HMF yet. The split-up becomes apparent at a redshift of $z=1.9$ and only steepens at $z=0.25$. In short, cosmic voids display a halo population similar to that of the entire simulation volume at cosmic dawn ($z=4.2$) and exert increasing impact as the underdensities deepen, with gravity progressively accumulating mass along the cosmic web. The same effect is also visible for the SMF, however, less pronounced, which could provide an additional constraint on cosmological parameters. 

  \item Not only do voids change the slope of the HMF, but the core density, as their global property, impacts the VHMFs as well. In that regard, voids with higher core density also allow for higher mass halos. This trend is also apparent for the void radius, where small voids tend to host a few more high-mass halos than large voids. This is consistent with the correlation of the two void-properties as only small voids up to $\approx 15$ Mpc/h show high core-densities (see \cite{schuster_nonlinear_2023}). However, the core density seems to be the primary void property that further limits the mass of hosted halos inside voids. Note that the ellipticity shows no significant impact on the VHMFs.  

  \item In a statistical sense, void galaxies show no inherent difference in baryonic processes, as the slope of the stellar-mass-halo-mass relation of void galaxies lies coherently within the contours of all galaxies within the simulation volume. This is a notable contrast to other numerical works \citep[i.e.,][]{habouzit:2020,alfaro:2020,rosasGuevara:2022,rodriguezMedrano:2024}, which do find differences in the star-formation behavior of void galaxies. While we find the assembly of the halos themselves to also be suppressed, this is not reflected in the stellar mass -- halo mass relationship, which is similar between void halos and halos in higher density regions, in agreement with observations \citep{douglas:2019}. If the lower halo masses of the void galaxies allow blue galaxies and spiral galaxies to survive at higher rates in voids, as some observations suggest \citep[e.g.][]{rojas:2004,rojas:2005}, this is not a conclusion that can be drawn from our statistical sample. This is due to the resolution of the simulation, which does not allow us to study galaxy properties beyond those presented in this work. Furthermore, and likewise due to resolution limits, our lowest halo masses are Milky Way-like galactic halos, which further limits statements about lower-mass galaxies.

  \item We find that cosmic voids have an impact on the SFR of galaxies at low redshifts ($z=0.25$). In voids, most high-mass galaxies exhibit quenched star formation. This effect is the strongest for galaxies within $0.25r_v$, as for masses $\rm{log(M_*[M_{\odot}])} \gtrsim 10.4 $ all galaxies have stopped their star-forming activity. For galaxies within $0.5r_v$, the number of star-forming galaxies also drops significantly at that mass, but does not reach zero. The fraction of star-forming galaxies after the cut-off mass increases with radius from the void center, as can be seen more clearly in the appendix in Fig.~\ref{fig:quenched_sf_vHMF}. The takeaway message is that void galaxies close to the void center are not able to continue star formation after reaching $\rm{log(M_*[M_{\odot}])} = 10.4 $ compared to the general galaxy population. However, this effect seems to appear only at low redshifts and until $z=2$, void galaxies follow the same star formation process as their non-void counterparts. More research should be done on whether this is driven by the lower availability of gas that could be depleted at $z=0.25$ or by other physical processes.

\end{itemize}

\begin{acknowledgements}
BAS acknowledges support by the grant agreements ANR-21-CE31-0019 / 490702358 from the French Agence Nationale de la Recherche / \emph{Deut\-sche For\-schungs\-ge\-mein\-schaft, DFG\/} through the LOCALIZATION project. BAS and KD acknowledge support by the COMPLEX project from the European Research Council (ERC) under the European Union’s Horizon 2020 research and innovation program grant agreement ERC-2019-AdG 882679. LCK acknowledges support by the DFG project nr. 516355818. This research was supported by the Excellence Cluster ORIGINS, which is funded by the Deutsche Forschungsgemeinschaft (DFG, German Research Foundation) under Germany's Excellence Strategy - EXC-2094-390783311. The Magneticum Pathfinder simulations were partially performed at the Leibniz-Rechenzentrum with CPU time assigned to the Project ``pr86re''. This work was supported by the DFG Cluster of Excellence ``Origin and Structure of the Universe''. We are especially grateful for the support of M. Petkova through the Computational Center for Particle and Astrophysics (C2PAP). 


\end{acknowledgements}

%
%

\bibliographystyle{aa}
\bibliography{bib}

\begin{appendix} 

\section{Impact of the Simulation Resolution on the HMF}\label{app_res}

\begin{figure}[htb]
   \centering
   \includegraphics[width=0.45\textwidth]{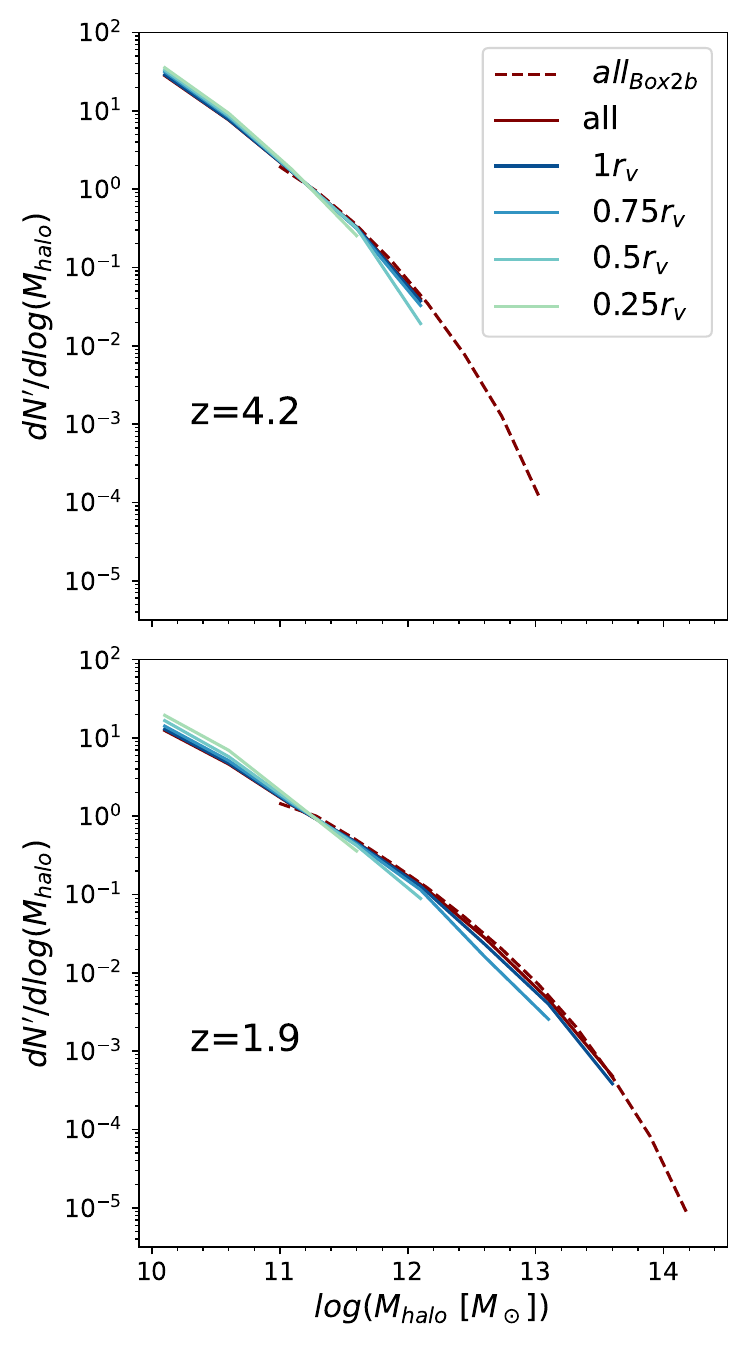}

   \caption{Using the higher resolution of \textit{Box3} and comparing the HMFs with the ones from \textit{Box2b} represented by the maroon-red dashed lines, one finds that they converge, suggesting that the HMFs do not depend on the resolution. Again, the split-up between the VHMFs is visible, colored in shades of blue to turquoise as in the figures before.  }
    \label{fig:resolution}%
\end{figure}

After investigating the different VHMFs and their evolution over time, we use \textit{Box3} to assess whether the findings from the previous section are resolution dependent. Since \textit{Box3} has available snapshots at redshifts of $z=1.9$ and higher, we use redshifts $z=4.2$ and $z=1.9$ to compare the results to \textit{Box2b}. Fig.~\ref{fig:resolution} depicts the set of VHMFs colored as previously, with the dashed line representing the general HMF in \textit{Box2b} at the corresponding redshifts. The split-up at a redshift of $z=4.2$ is not very distinct, however, faintly visible, hinting at the split-up that will evolve and show at a redshift of $z=1.9$. Compared to the VHMFs in \textit{Box2b}, the $0.25r_v$ and $0.5r_v$ VHMFs show a distinguishable split-up already at a redshift of $z=4.2$. 
However, in the bottom panel, the split-up appears to be less pronounced than in Fig.~\ref{fig:vHMF_evo}.
This is caused by an interplay of the different resolutions and the significantly larger volume of Box2b.  To account for the larger mass range of Box3 at the low-mass end, we set the normalization described in section 3.2 to the total number of halos in the mass range that is covered by both boxes, excluding the smaller halos that are resolved in Box3 but not in Box2b. This drives up the low-mass end for this simulation, amplifying the discrepancy between the different VHMFs. On the high-mass end, Box3 intrinsically does not reach the same halo masses that Box2b does, presumably due to the smaller box size suppressing the large-scale density fluctuations that give rise to the most massive halos. 

However, accounting for these statistical differences, the general HMF of \textit{Box2b} and \textit{Box3} converges, proving that the resolution of the simulation volume does not, in fact, influence the HMF. Additionally, the split-up of the VHMFs we observed in Fig.~\ref{fig:vHMF_evo} is also visible, however, with a more pronounced low-mass end showing the overabundance of low-mass halos within voids as described earlier.

\section{More on the Dependence of the VHMFs on Void Properties}

\begin{figure*}[htb]
   \centering
   \includegraphics[width=0.98\textwidth]{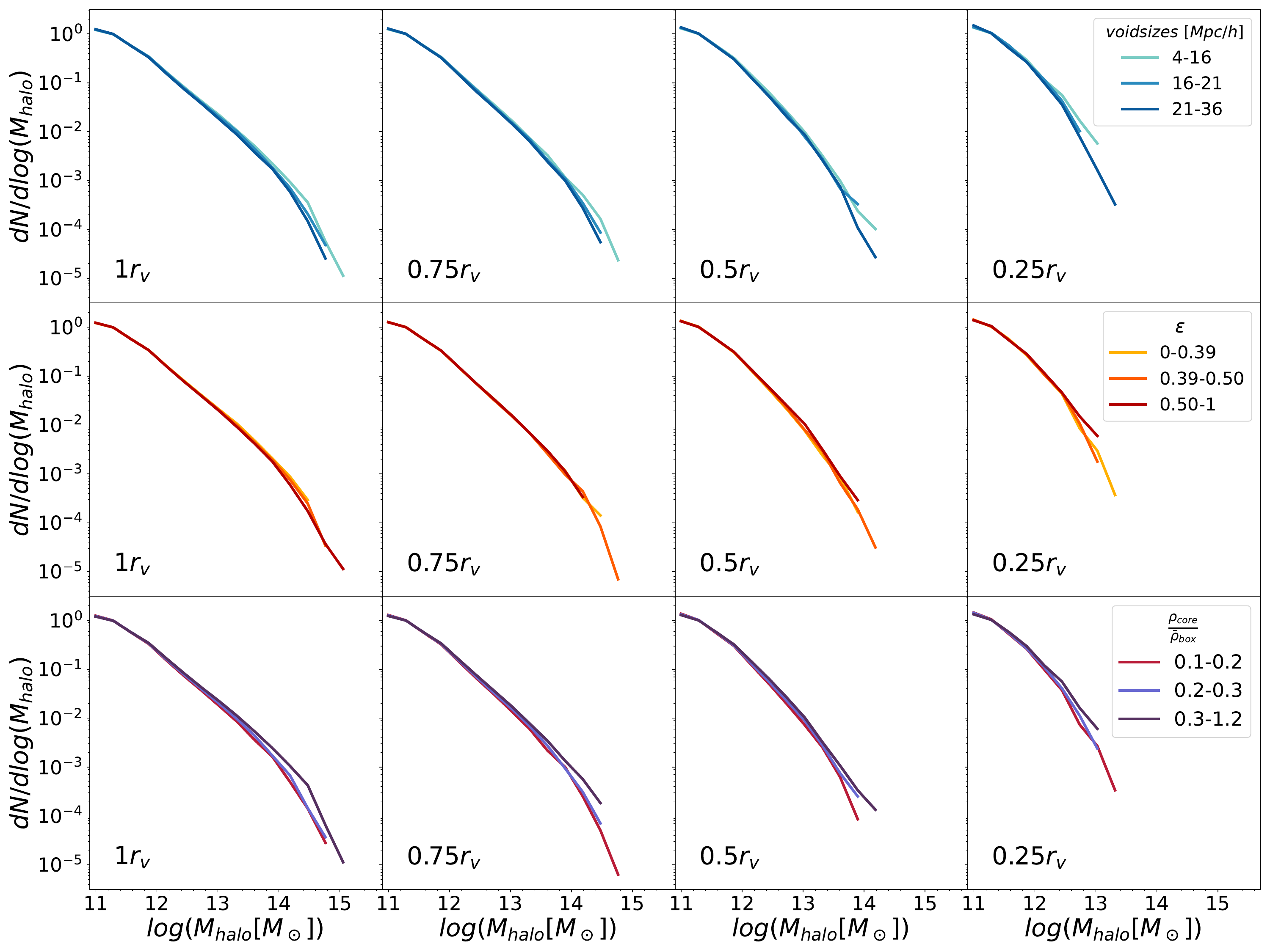}

   \caption{Investigating different void properties and their impact on the VHMFs. The top row in blue shows the VHMFs depending on the size of the voids using three equally sized bins, where the light blue line includes only halos that lie within voids with sizes from 4 to 13 Mpc/h, in a darker blue voids with sizes 13-18 Mc/h are evaluated, and in the darkest blue sizes 18-36 Mpc/h. Colored in orange, the middle row splits the VHMFs depending on the ellipticity of the voids, ranging from low to high ellipticity in light orange to red, respectively. In the bottom row the density contrast $\rho_{core}/\bar{\rho}_{box}$is analyzed, with $\rho_{box}$ describing the mean density in the simulation volume and $\rho_{core}$ is the density at the void center, from dark red to deep purple corresponding to a range from low to high density contrast. In each row from left to right is the } 
              \label{fig:split}%
\end{figure*}

To see whether the split-up is only visible in the outermost percentiles or if this is a general trend we proceeded as follows: The halo catalog was first filtered according to the position of the halos inside the void corresponding to the respective VHMF (e.g., the $1r_v$ -VHMF includes all halos that lie within the radius of the void), and then the subsets were sorted by the property of interest and divided into three equal-count bins. 

The first row of Fig.~\ref{fig:split} shows the set of VHMFs from left to right, each split into three bins by void size. The light blue line includes all halos that reside in small voids with sizes ranging from $4-16 Mpc/h$, the darker blue contains void-sizes of $16-21 Mpc/h$, and the darkest blue considers the largest voids with a range of $21-36 Mpc/h$. The impact of the void size on the VHMFs seems to be negligible; however, there is a hint that the larger voids host less massive halos than the small ones. The fact that the larger voids arise from steeper initial density perturbations could explain why they evolve to host less massive halos. In the second row of Fig. ~\ref{fig:split}, we investigate the impact of ellipticity on the VHMFs. The red line depicts the most elliptical voids with an ellipticity of $\epsilon = 0.5-1$, the orange line contains voids with an ellipticity of $\epsilon = 0.39-0.5$, and the most spherical voids with $\epsilon = 0-39$ are colored light orange. The three VHMFs clearly converge, justifying the spherical approximation of the voids in the previous calculations. The last row of Fig.~\ref{fig:split} visualizes the impact of the core density on the VHMFs, showing the msot pronounced effect especially between the very low-density environments with density contrasts of $\rho_{core}/\bar{\rho}_{box} = 0.1-0.2$ (raspberry red) compared to high core densities of $\rho_{core}/\bar{\rho}_{box} = 0.3-1.2$ (blue violet).

\section{Environment Dependent Star-Formation in Voids}
\begin{figure}
    \centering
    \includegraphics[width=0.98\columnwidth]{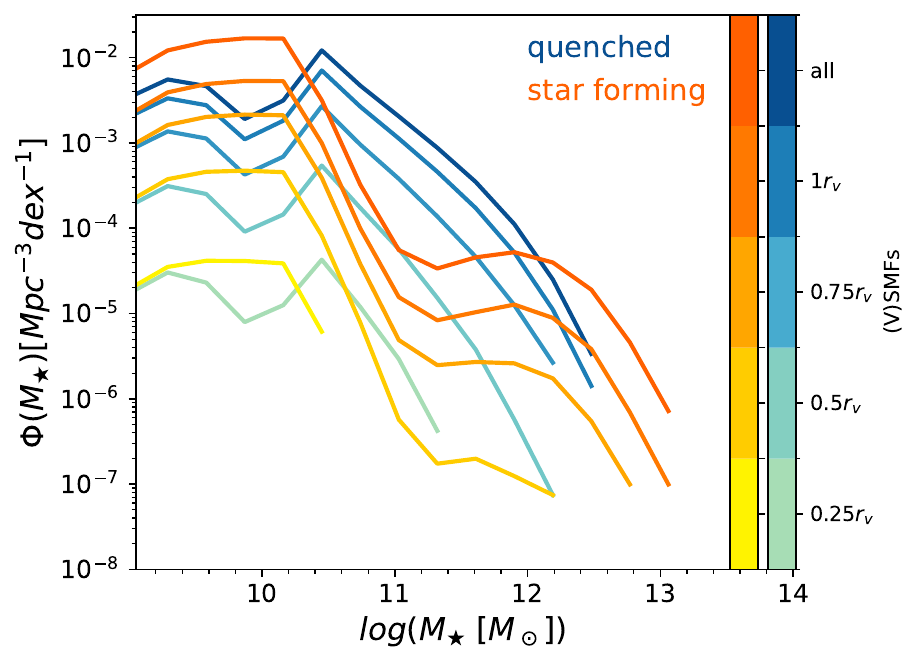}
    \caption{(V)SMFs for quenched (blue-turquoise) and star-forming (orange-yellow) galaxies, separately, for the set of four spherical shells $1,\ 0.75,\ 0.5,\ 0,25 r_v$ from the void center. The lines that should be compared to each other are depicted as neighbors in the two colorbars on the right.}
    \label{fig:quenched_sf_vHMF}
\end{figure}

Figure ~\ref{fig:quenched_sf_vHMF} highlights the results of the rightmost panel in Fig.~\ref{fig:sfr}. It shows the set of VSMFs calculated for star-forming (orange-yellow) and quenched galaxies (blue-turquoise) with a volume norm for better visibility. The most drastic result is evident when comparing the $0.25r_v$ VSMF for the star-forming (yellow) and quenched (turquoise) galaxies. The star-forming galaxies can only be found up to masses of $\approx 10^{10.4} M_{\odot}$ while the VSMF for quenched galaxies reaches masses up to $\approx 10^{11.25} M_{\odot}$. This trend is not visible from $0.75r_v$ onward, including the general galaxy population. Therefore, the most underdense regions inside voids not only act upon galaxies as a mass limiting factor, but also limit star-forming activities to smaller galaxies, while more massive galaxies only seem to be able to continue star-formation in more dense environments. The nature of this effect has yet to be studied.    

\end{appendix}

\end{document}